\documentclass[preprint,12pt]{revtex4}
\usepackage{graphicx}
\usepackage{subfigure}
\usepackage{hyperref}
\usepackage{listings}
\usepackage{exscale}
\usepackage{array}
\usepackage{multirow}
\usepackage{color}
\usepackage{amsmath,amssymb}
\usepackage{booktabs}

\newcommand*{\vcenteredhbox}[1]{\begingroup
\setbox0=\hbox{#1}\parbox{\wd0}{\box0}\endgroup}

\begin{document}
\lstset{language=C++}

\author{M.~Bernaschi$^\dagger$, M.~Lulli$^\diamond$, M. Sbragaglia$^\diamond$}
  \address{$^\dagger$Istituto per le Applicazioni del Calcolo CNR, Via dei Taurini 9, 00185 Rome, Italy.\\
  $^\diamond$Department of Physics and INFN, University of Tor Vergata, Via della Ricerca Scientifica 1, 00133 Rome, Italy
}

\title{GPU Based Detection of Topological Changes in Voronoi Diagrams}

\begin{abstract}
  The Voronoi diagrams are an important tool having theoretical and practical
  applications in a large number of fields. We present a new
  procedure, implemented as a set of CUDA kernels, which detects, in a
  general and efficient way, topological changes in case of
  dynamic Voronoi diagrams whose generating points move in time.
  The solution that we provide has been originally
  developed to identify plastic events during simulations of
  soft-glassy materials based on a Lattice Boltzmann model with
  frustrated-short range attractive and mid/long-range
  repulsive-interactions. Along with the description of our approach,
  we present also some preliminary physics results.
\end{abstract}

\maketitle
\section{Introduction}

Identifying entities, characterizing them according to a minimal set of their most significant features and studying their dynamics looking for interesting patterns, is a class of problems found in many different contexts, from soft-matter physics to computational geometry. In the present paper we describe a combination of algorithms and parallel computing techniques that solve an important instance of such problems by using as computing platform a GPU programmed according to the CUDA model. The work has been originally motivated by the need of detecting plastic rearrangements of emulsion droplets \cite{Larson,Goyon08} in numerical simulations based on a recently introduced lattice Boltzmann model with frustrated interactions \cite{ourJCP}. The model allows for the mesoscopic description of binary fluids, specifically a collection of droplets dispersed in a continuous phase. The coalescence of these droplets is inhibited by the (positive) disjoining pressure which develops in the thin films separating two approaching interfaces. The model gives direct access to the hydrodynamical variables, i.e. density and velocity fields, as well as the local (in time and space) stress tensor in the system. Thus, it is extremely useful to characterize the relationship between the droplets dynamics, their plastic rearrangements, and the stress fluctuations. Once the droplets are stabilized, different packing fractions and polydispersities can be obtained and simulated under different load conditions \cite{Soft14,Soft15,RoughnessEpl2016}.

From a formal viewpoint, the problem corresponds to the computation of the Voronoi diagram of moving points. The Voronoi diagram of a set, $S$, of $n$ objects in a space $E$ is a subdvision of this space into regions, so that all points within a region have the same nearest neighbor in $S$ according to a given distance measure $d$ \cite{voronoi}. The dual graph of the Voronoi diagram is the Delaunay triangulation of the input set, a planar graph that connects two sites by an edge whenever there exists a circle that contains those two sites and no other site \cite{delaunay}. Topological changes in a Voronoi diagram correspond to edge flips in the Delaunay triangulation. For our model, a topological change is a signal that a plastic event happened during the simulation. Although there is a significant amount of theoretical work providing techniques to minimize the number of operations required to update Voronoi diagrams (starting on seminal works like \cite{DynVor}), the effort to provide high performance parallel implementations for the solution of the problem has been much more limited \cite{DTGPU}. We believe that the procedure we developed is powerful and general enough to be useful for other applications. To support the idea, we provide, along with the source code of the CUDA implementation, a sample code that shows how to use the procedure to study the dynamics of {\em bubbles} that form according to a simple rule applied to the value of a random field defined on a two-dimensional mesh.

The rest of the paper is organized as follows. Section \ref{sec:cenvor} describes, mainly through examples, applications of the {\em centroidal} Voronoi tessellation that is the case in which the generating points of the Voronoi diagrams are, in a broad sense, the {\em  centers of mass} of different types of entities. Section \ref{sec:pro} describes our CUDA procedure to detect topological changes in Voronoi diagrams, that will be available from \verb|http://twin.iac.rm.cnr.it/dynvorcuda.tgz| as soon as we complete the user documentation. Section \ref{sec:lbglassy} reports some preliminary results obtained using the {\em on-line} detection of plastic events during the simulation of our lattice Boltzmann model of emulsions \cite{Soft14,Soft15,RoughnessEpl2016}. Finally, Section \ref{sec:conc} concludes the work summarizing the results and providing future directions of activity.

\section{Centroidal Voronoi Diagrams: Definition and Applications \label{sec:cenvor}}
Given a region $U \subseteq \mathbb{R}^N $ and a density function $\rho$, defined in U, the {\em centroid} $\mathbf{z}^*$ of $U$ is defined by
$$
\mathbf{z}^*=\frac{\int_U \mathbf{y}\rho(\mathbf{y}) d\mathbf{y}}{\int_U \rho(\mathbf{y}) d\mathbf{y}}
$$
The Voronoi regions $\{\widehat{U}_i\}$, given $k$ points $\mathbf{z}_i, i=1,....,k$ in $U$ are defined as:
$$
\widehat{U}_i=\{\mathbf{x} \in U \mid |\mathbf{x}-\mathbf{z}_i| < |\mathbf{x}-\mathbf{z}_j|~{\rm  for}~j=1,...,k, j\ne i\}
$$
The {\em centroidal Voronoi tessellation} corresponds to the situation
where
$$
\mathbf{z}_i=\mathbf{z}_i^*,~i=1,...,k
$$
There are several applications of centroidal Voronoi tessellations in fields like data compression, numerical analysis, biology, statistics and operations research \cite{CentVoronoi}. In the present Section we briefly present a few examples to support the idea that our GPU-based procedure for the detection of topological changes in centroidal Voronoi diagrams can be useful in the study of a wide range of important problems.

\subsection{Soft-Matter Physics}

As a first example, we recall the importance of centroidal Voronoi diagrams in the understanding of the yielding transitions and flow properties of soft-glassy materials. These include a variety of complex fluids such as emulsions, foams, gels, and many others \cite{Larson}. These materials pose important challenges from the fundamental point of view of out-of-equilibrium statistical mechanics, hence they have been the object of intense scrutiny in the recent years \cite{Bouzid}. When the packing fraction of their elementary constituents (droplets for emulsions, bubbles for foams, etc) exceeds a critical value, the system undergoes a {\it jamming} transition and a yield stress emerges, below which the material responds elastically and above which it flows like a non-Newtonian fluid. These complex flow properties are even more complicated in case of confining geometries \cite{Goyon08,Mansard14}, where non-local effects and spatial cooperativity \cite{Goyon08} have been invoked to explain the observed flow properties. A number of theoretical frameworks has been developed to take into account these non-local effects (see \cite{Bouzid} and references therein). The basic phenomenology can be captured by relating the flow properties to the rate of plastic rearrangements in the system \cite{KEP}. In the case of emulsions, plastic rearrangements are identified with $T1$ topological events, in which neighbouring bubbles switching occurs. Given the centers of mass of the bubbles, centroidal Voronoi diagrams are then crucial to reveal the properties of the underlying topology and study plastic rearrangements: by following the evolution of the Voronoi diagrams of four bubbles, one can exactly identify the $T1$ plastic events when the edge length of a Voronoi cell goes to zero, or equivalently, when the diagonal edge of the Delaunay triangulation flips. This has been indeed used by various works in the literature \cite{Mansard13,Soft14,Soft15,ourJFM}. Accessing centroidal Voronoi diagrams with efficiency and flexibility, especially in those situations where the soft-glassy materials flow in confined geometries with complex boundary conditions \cite{RoughnessEpl2016,Mansard14}, is therefore an important issue which raises many challenges ahead.

\subsection{Clustering Analysis}
Centroidal Voronoi diagrams play an important role in clustering analysis, a general tool used in many disciplines including (but not limited to) pattern recognition, computer graphics, combinatorial chemistry. Given a set $ S $ of objects, the goal is to split the set into $k$ disjoint subsets, that determine the best classification of the objects according to one or more criteria that distinguish among them. The technique is often called $k-$clustering analysis \cite{kmeans}. For instance, in combinatorial chemistry, $k-$clustering analysis may be used in compound selection where the similarity criteria could be either compound components and/or their structure \cite{clark}. Here for the purpose of illustrating the connection between centroidal Voronoi tessellation and the optimal $k-$clustering, we consider a very simplified situation where $S$ contains $m$ points in $\mathbb{R}^N$. Given a cluster $U_i$ of $S$ with $n$ points, the cluster is formally represented by a value $\mathbf{c}_i$.  The {\em squared error} with respect to $\mathbf{c}_i$ is given by
$$
\mbox{SE}(U_i) = \sum_{\mathbf{x}_j\in U}|\mathbf{x}_j-\mathbf{c}_i|^2
$$
and for a $k-$clustering $\{U\}_{i=1}^k$ (a tessellation of $S$ into $k$ disjoint subsets), the so-called Within-Cluster Sum of Squares (WCSS) \cite{kmvoronoi1} corresponds to
$$
\mbox{WCSS}(S) = \sum_{i=1}^k \mbox{SE}(U_i) = \sum_{i=1}^k\sum_{\mathbf{x}_j\in U_i}|\mathbf{x}_j-\mathbf{c}_i|^2.
$$
By differentiating the expression of the WCSS with respect to $\mathbf{c}_i$, it is easy
to find that the optimal $k-$clustering, having the minimum WCSS, occurs when $\mathbf{c}_i$ is represented by the arithmetic mean
$$
\overline{\mathbf{x}}_i=\frac{1}{n}\sum_{\mathbf{x}_j\in U}\mathbf{x}_j
$$
corresponding to the centroid of $U_i$, i.e., the optimal $k-$clustering is a centroidal Voronoi diagram \cite{kmvoronoi2} or, in other words, $\{U\}_{i=1}^k$ is the Voronoi partition of $S$ having $\{\overline{\mathbf{x}}\}_{i=1}^k$ as the generators of the partition.

\subsection{Optimal Placement of Resources}

As another example, we describe the problem of optimal placement of $k-$mailboxes (or any other set of shared resources, like schools, distribution centers, {\em etc.}) in a city or neighborhood so as to make them most convenient for users. If we assume that:
\begin{itemize}
\item users prefer the mailbox nearest to their home;
\item the cost (to the user) of using a mailbox increases as the
  distance from the user's home to the mailbox;
\item the total cost for all users is given by the distance to the
  nearest mailbox averaged over all users;
\end{itemize}
and, finally, that
\begin{itemize}
\item the optimal placement minimizes the total cost to the users,
\end{itemize}
then the total cost function corresponds to:
\begin{equation}
\sum_{i=1}^k \int_{U_i}
|\mathbf{x}-\mathbf{x}_i|^2\phi(\mathbf{x})d\mathbf{x}
\label{eq:oppl}
\end{equation}
where $\phi$ is the population density. It is apparent that a placement defined as optimal according to \ref{eq:oppl} is at the centroids of a centroidal Voronoi tessellation of the region.

\subsection{Distribution of Animals}

As a last example, we mention the behaviour of animals when they stake out a territory. When the animals settle into a territory one or a few at a time, very often the distribution of Voronoi generators is similar to that of centers of circles (with approximately fixed radius) in a random circle packing problem. However, when all the animals settle at the same time, {\em i.e.}, in a {\em synchronous} way, the distribution of Voronoi centers is very likely that for a centroidal Voronoi tessellation. In \cite{CentVoronoi} are summarized the results of a controlled experiment carried out in a large outdoor pool with a uniform sandy bottom. A bunch of mouthbreeder fishes were introduced in the pool. Each male of the mouthbreeder fish excavates a breeding pit in the sand by spitting sand away from the pit center. Since the fish tend to stay as far away as possible from their neighbors, after an adjustment process, each fish had its spitting location that coincided with a centroidal Voronoi tessellation. In that case, the generators of the tessellation moved towards the centroids until a steady state configuration was established.

\section{Detection of  topological changes in Voronoi
  diagrams \label{sec:pro}}In the present Section we describe the
steps of our CUDA procedure for the detection of topological changes
in Voronoi diagrams. Although for the description we refer, at times, to our Lattice Boltzmann model of soft-matter, the procedure can be applied with very few changes to any other Voronoi diagram of moving points.
\subsection{Identification of bubbles }
\begin{figure}[!h]
\centering
\includegraphics[scale=0.5]{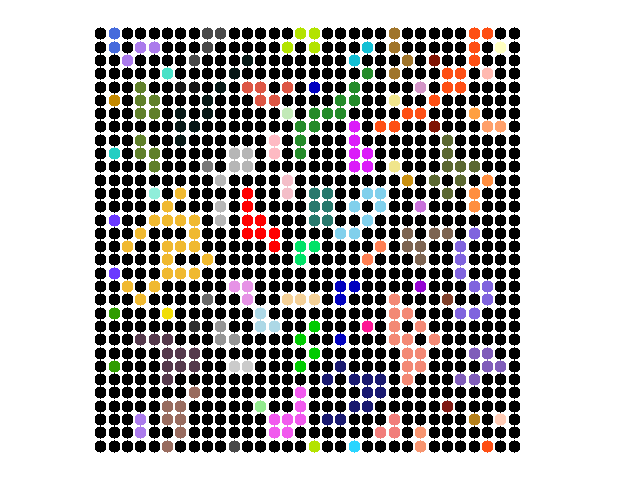}
\caption{Bubbles on a 2D lattice. Black points do not belong to
  {\em bubbles} because the  value of the density is higher with
  respect to the threshold.}
\label{fig:bubbles}
\end{figure}

Hereafter with the term {\em ``bubble''} we mean a set of points of a mesh
such that the value of a field $\rho$ ({\em e.g.}, the density) is below a
predefined threshold for all points in the set and each point of the set can be reached starting
from any point of the same set by moving along the directions of the Lattice Boltzmann model under study.
This means that two bubbles never share points otherwise they are
fused in a single bubble. Figure \ref{fig:bubbles} shows sample
bubbles (shown as colored points) in a $2d$ lattice. The
identification of such type of bubbles resembles remarkably the
problem of {\em cluster-labelling} whose solution is required as a part of
the Swendsen-Wang algorithm for the flip of spin-clusters \cite{SW}. A cluster-labelling
procedure assigns unique labels to each distinct cluster of lattice
points. A cluster is a set of lattice points that are connected to each
other according to the interaction rules of the spin system and
 that have the same value of the spin for all the points in the
 cluster. To exploit the apparent analogy between our problem of {\em
   bubbles} identification and the {\em cluster-labelling} problem, we
 introduce a dummy variable for each point of the lattice whose value
 is $1$ if $\rho$ is below the threshold value that defines a point as
 belonging to a bubble and $-1$ otherwise. In this way we have a $2d$
 Ising-like  spin system ({\em i.e.}, with two possibile values) and
 we can leverage existing efficient GPU algorithms for
 cluster-labelling \cite{Labeling}. Once each {\em bubble} has been
 identified with a unique label, we determine its center of mass. To
 that purpose we need to count the number of lattice points in each
 bubble. This can be done either with a simple kernel that uses atomic
 operations or by building a histogram where the bins correspond to the labels
 that identify the bubbles. The latter approach can be effectively
 implemented by using a set of primitives available in the CUDA {\em thrust}
 library:
\begin{enumerate}
\item sort data to bring equal elements together;
\item find the end of each bin of values;
\item compute the histogram by taking differences of the cumulative histogram.
\end{enumerate}
To obtain the coordinates of the center of mass of each bubble we need
to sum, separately, the coordinates of the lattice points that form
the bubble and divide by the size of the bubble ({\em i.e.,} the
number of lattice points). Also in this case it is possible to resort either
to atomic operations or to apply a reduction operation to the
arrays that contain the coordinates ($x$ and $y$ in two dimensions) of the points
belonging to each bubble.

\subsection{Digital Voronoi diagram and Delaunay Triangulation}
We build the Voronoi diagram using the centers of mass
$\{\mathbf{z}_i^*\}$ of the bubbles as generating points.
We want to exploit at its best the fact that the input density function
$\phi(\mathbf{x})$ is defined on a lattice. As a consequence, we cannot directly aim at obtaining
a Voronoi tessellation which is compliant with its continuous
definition. Nonetheless, we can define the so-called \emph{digital Voronoi}
tessellation \cite{Rong:2008, Qi:2012, maljovecdelaunay} which can be used to define the dual Delaunay triangulation.

Since the Delaunay triangulation only involves topological properties, one can compute the \emph{continuous} Voronoi tessellation starting from the circumcenters of the triangles, which, in general, is going to differ from its digital counter part. Hence, the lattice acts as a natural cutoff for distances. Such a cutoff is expected to induce differences between the digital and the continuous case for those configurations which are close to degeneracy.

Degenerate configurations occur any time a vertex of the Voronoi
diagram is shared by more than three centers of mass, or, from the
Delaunay point of view, when, given four or more points, it is
impossible to choose the diagonal bonds \cite{Dillencourt1993}. From
the geometrical point of view, such configurations are originated
whenever four or more centers of mass lie extremely close to the same
circumference. As a matter of fact, topological changes in the Voronoi tessellation necessarily involve such degenerate configurations.

Since in a discrete setting topological properties are more
robust than metric ones, we choose to detect plastic events from
changes in the Delaunay triangulation of the centers of mass. We also
choose to look at the triangulation as a graph rather than a set of
triangles. Then, we are naturally led to employ the concept of
\emph{adjacency matrix} in order to describe the triangulation. At the
very end of the process, we are able to determine whether or not a
topological change happened by comparing the adjacency matrices once a suitable isomorphism between the labels of the bubbles of two different configurations is given.

In the following we describe how to perform a digital Voronoi tessellation and then obtain its dual Delaunay triangulation in the form of an adjacency matrix.

\subsubsection{Digital Voronoi tessellation}
\begin{figure}[!t]
\centering
\includegraphics[scale=0.63]{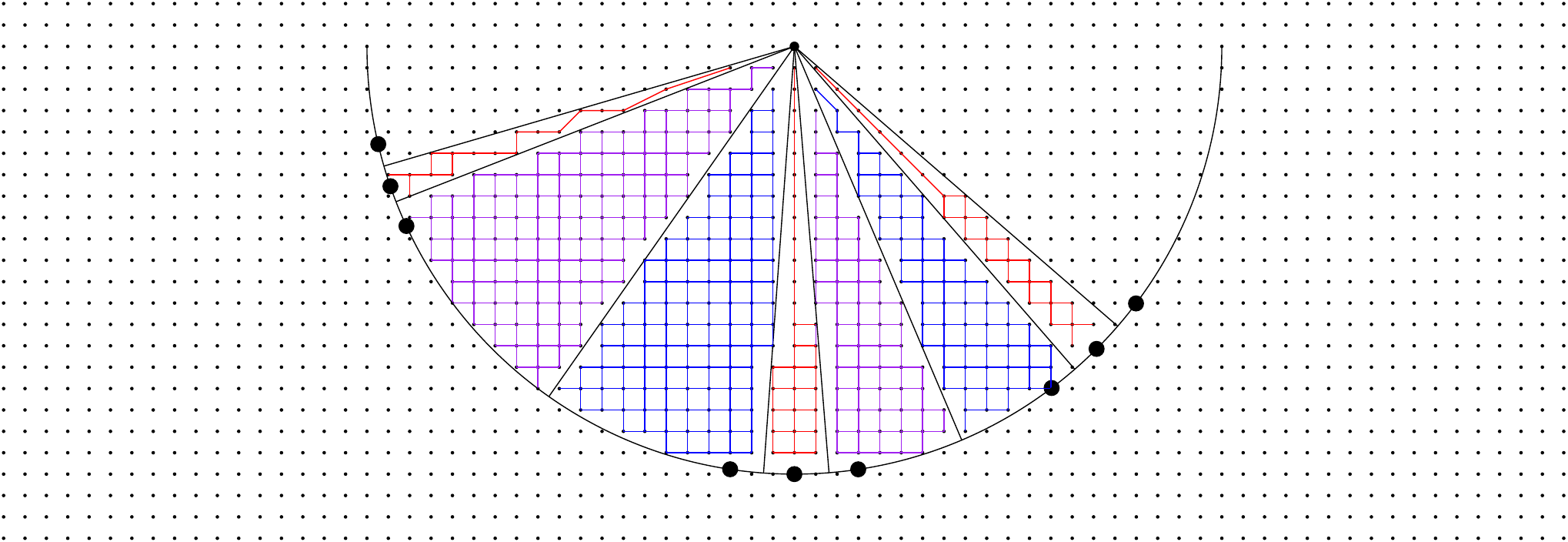}
\caption{Example of colouring defects for the digital Voronoi diagram with
  Euclidean distance on a lattice: the refernce centers of mass are placed
  along a cirumference, \emph{i.e.} in a degenerate configuration, and they are
  placed at a very small angle with respect to its center, thus producing the
  colouring pathologies. In the top-left red subdivision, it is
    possible to see, close to the center, the so-called
    \emph{islands}, whereas in the bottom-center red subdivision it is
    possibile to see the so-called \emph{prolongued vertex}}
\label{fig:degeneracy}
\end{figure}

In order to define a Voronoi tessellation, let us start by looking at a pathological configuration. In
Figure \ref{fig:degeneracy}, we report an example in which a number of
centers of mass is placed extremely close to a common circumference with no other
point inside of it.

A pretty simple {\em parallel} solution for the digital Voronoi tessellation with Euclidean distance is the following: given the labels of the centers of mass, one associates a thread to each lattice
site; each thread computes its distance from all the centers of mass finding
the closest one and thus \emph{coloring}\footnote{From now on we will use the terms \emph{labels} and \emph{colours} as equivalent, since the labels of the bubbles will be used as colours for the digital Voronoi tesselation.} itself with the label of the
nearest center of mass. The task can be optimized by calculating just the
squared distance, \emph{i.e.}, avoiding the square root.

However, as it was already
reported in \cite{Rong:2008, Qi:2012, maljovecdelaunay}, the definition of digital Voronoi diagrams based
on the Euclidean distance leads to pathologies. This can be seen in
Figure \ref{fig:degeneracy}: for a Voronoi cell with a sharp angle, it is
possible to have isolated points (referred to as \emph{islands} in
\cite{Rong:2008, Qi:2012, maljovecdelaunay}), \emph{i.e.}, points with the first eight neighbours coloured
differently, leading to a non simply-connected Voronoi cell on the
lattice. We also name \emph{prolongued vertices} those colourings for which only a single line of points is coloured running from side to side in the stencil. However, the
concept of pathology is closely related to the algorithm which is adopted in
order to obtain the Delaunay triangulation starting from the digital Voronoi
tessellation. As we show in the next subsection, the set of rules we
choose, which differs from that of \cite{Rong:2008, Qi:2012, maljovecdelaunay}, automatically handles Euclidean colouring pathologies.

Another interesting aspect arises for what concerns constraints on the digital
Voronoi diagrams: it is possible to manage easily the presence of
obstacles in the
domain by defining a supplementary binary field which takes the value $1$ on
those sites which are considered obstacles and $0$ otherwise. While colouring
such points, we can set their colour to be a fixed large number,
\texttt{MAXLINK}\footnote{In our case \texttt{MAXLINK=0x77777777}}, which is
certainly larger than the total number of bubbles. By doing so, we define
regions of the lattice with undefined colour which automatically will not be considered for
the Delaunay triangulation. It is important to highlight that in this way complex
geometries are naturally handled. We will come back again to this point in the
next subsection.

\subsubsection{Delaunay triangulation}
Once we are given a digital Voronoi tessellation, we can determine the
adjacency matrix for the Delaunay triangulation just by analyzing the vertices
of the Voronoi cells. Indeed, at the vertices it is possible to determine which
colours are in contact in a non-ambiguous way. This choice has a two-fold
reason: on one hand the number of vertex sites is much smaller than the
number of edge sites, so most of the warps \footnote{a {\em warp} is a group of 32 CUDA threads} do not branch; on
the other hand, for those configurations that are very close to degeneracy, the
topological information can be retrieved only at the vertices. Still, we want to
emphasize that the discrete nature of the lattice introduces a bias with respect
to the continuous case.

\begin{figure}[!t]
\centering
\includegraphics[scale=1.0]{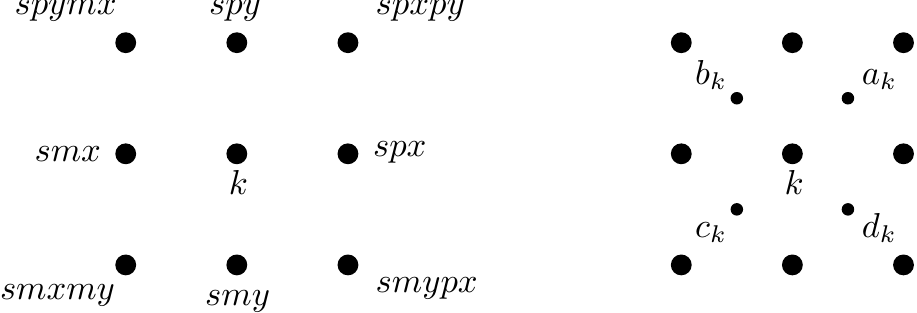}
\caption{Bulk stencil definition: on the left panel the labels for the frame
  points are given, whereas on the right panel there are those for the points of the inner-frame.}
\label{fig:bulkFrame}
\end{figure}

We define a point to be a vertex according to the properties of its
\emph{frame}. In the bulk, the frame of a given lattice point $k$ is
the set of its eight nearest neighbours (defined according to the standard
  LB directions in $2d$) that we name as reported in Figure
\ref{fig:bulkFrame}. On the domain boundaries, the definition changes accordingly: the
frame is given by all nearest neighbours that do not lie outside the
domain. In order to treat diagonals, we also need to consider other four points
$\{a_k, b_k, c_k, d_k\}$ which we call the \emph{inner-frame}.
We refer to the set of points in the frame, inner-frame and $k$ as \emph{stencil}.

The strategy we propose is to differentiate nodes that lie at the boundary
between two Voronoi cells from those that lie at the vertices by counting the
number of \emph{gaps} along the frame, \emph{i.e.}, how many times two nodes
along the frame belong to two different Voronoi cells.

\begin{figure}[!h]
\centering
\includegraphics[scale=1.0]{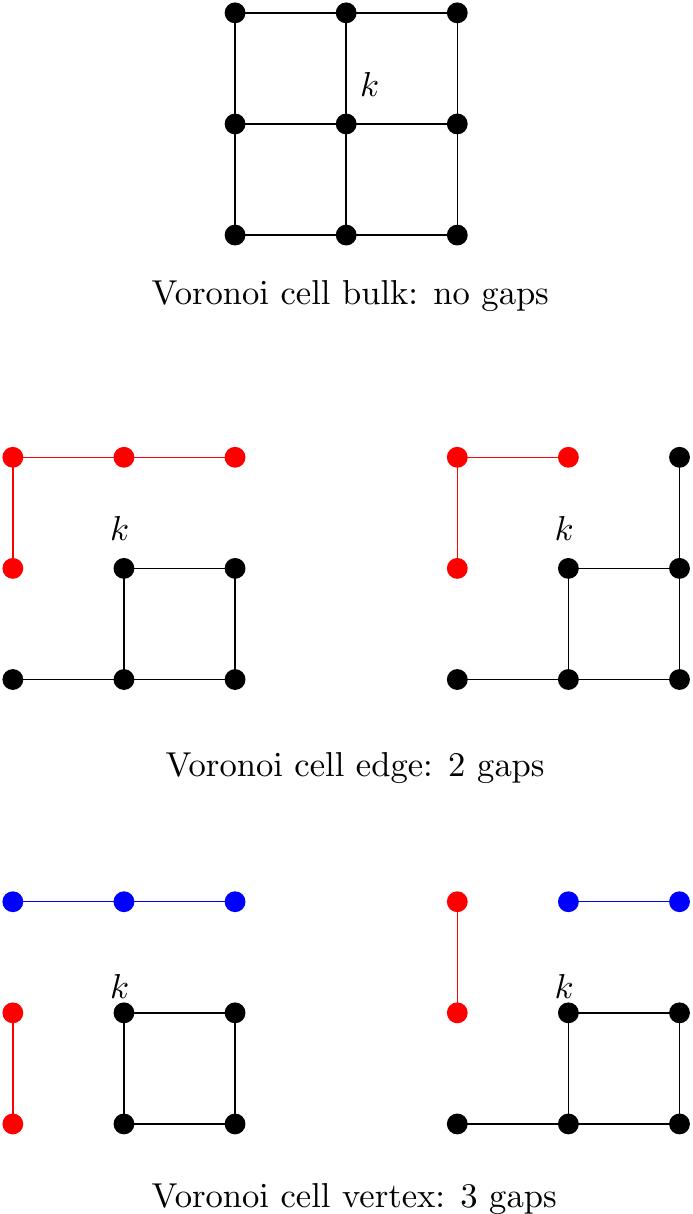}
\caption{Top panel: $k$ is a bulk site. Second row: the point $k$ lies along an edge between two
  Voronoi cells: there are only two colour gaps along the frame. Third row: the point $k$ lies at the vertex of a Voronoi cell where three colour
  gaps can be found along the frame.}
\label{fig:edgeVertexExamples}
\end{figure}

With such a definition, bulk points possess frames with no
gaps, whereas lattice points at the boundary between two Voronoi cells
must have two gaps in their frames. Vertex points are those with three or more
gaps along the frame. In Figure \ref{fig:edgeVertexExamples} we report some
examples of colour configurations for bulk, boundary and vertex sites.

\begin{figure}[!t]
\centering
\includegraphics[scale=1.0]{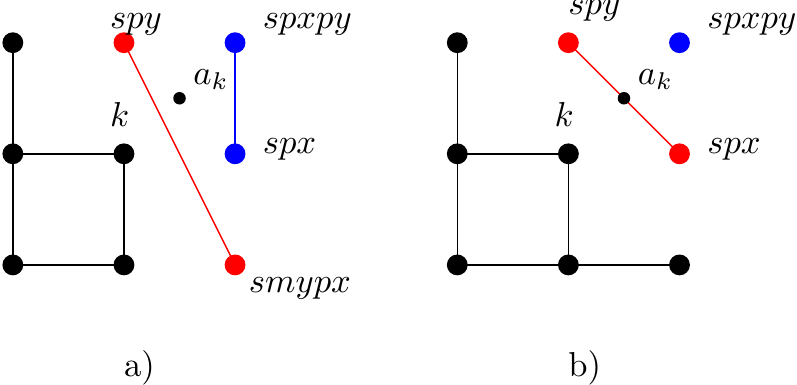}
\caption{Frame pathologies. \textbf{First-nearest neighbours} rules: a)
  panel is a pathological colouring for $spx$ whereas the b) panel is
  not. \textbf{Second-nearest neighbours} rules: both a) and b) represent
  pathological colourings for $spxpy$.}
\label{fig:framePathologies}
\end{figure}

Once the lattice sites on Voronoi vertices have been recognized, it is
necessary to
choose which of the colours along the frame are to be considered neighbours of
the colour carried by the site $k$, \emph{i.e.}, which of the centers of mass
$\{\mathbf{z}^*\}$ are to be considered neighbours in the Delaunay
triangulation. We now describe the construction of the adjacency
matrix once a vertex point has been recognized. It is necessary to distinguish between
nearest neighbours ($\{spx, spy, smx, smy\}$), and second
nearest-neighbours ($\{spxpy, spymx, smxmy, smypx\}$). Moreover, we introduce
the notation for the colour of a given point $C(X)$, where $X$ indicates all
the labels within the considered stencil. As a first step, all entries of
the adjacency matrix are set equal to \texttt{MAXLINK}. Then, the set of rules needed to determine the Delaunay triangulation neighbours of a bubble, identified by the colour/label $C(k)$, at the vertex of its Voronoi cell is the following:
\begin{itemize}
\item \textbf{first nearest-neighbours}. Let us consider the case of the point
  $spx$: $C(spx)$ is a neighbour of $C(k)$ if
  \begin{enumerate}
  \item $C(spx) \neq \texttt{MAXLINK}$.
  \item $C(spy) \neq C(smypx)$,
  \item $C(spx) \neq C(k)$,
  \end{enumerate}
  The first and third conditions are readily understood whereas the second one handles the
  long diagonal pathologic colouring reported in Figure \ref{fig:framePathologies}.
\item \textbf{second nearest-neighbours}. Let us consider the case of the point
  $spxpy$: $C(spx)$ is a neighbour of $C(k)$ if
  \begin{enumerate}
  \item $C(spxpy) \neq \texttt{MAXLINK}$.
  \item $C(spxpy) \neq C(k)$,
  \item $C(spy) \neq C(smypx)$,
  \item $C(spy) \neq C(spx)$,
  \item $C(a_k) = C(spxpy)$ or $C(a_k) = C(k)$,
  \end{enumerate}
  While the fourth condition handles the short diagonal pathologic colouring,
  the third condition handles the long diagonal one, see Figure
  \ref{fig:framePathologies}. The fifth condition decides whether or not the two different colours $C(spxpy)$ and $C(k)$ are to be linked in the Delaunay triangulation.
\end{itemize}
Let us briefly comment upon the fifth condition for second
nearest-neighbours: it is clear that this condition is somewhat
arbitrary. Indeed, this choice introduces a cutoff. However, we expect such a
choice to influence only those points very close to a degenerate configuration, which is, in the general case, the
most important issue on which different algorithms may produce a
different result due to the features of floating point arithmetics
({\em i.e.}, non-associativity).

\begin{figure}[!t]
\centering
\includegraphics[scale=1.0]{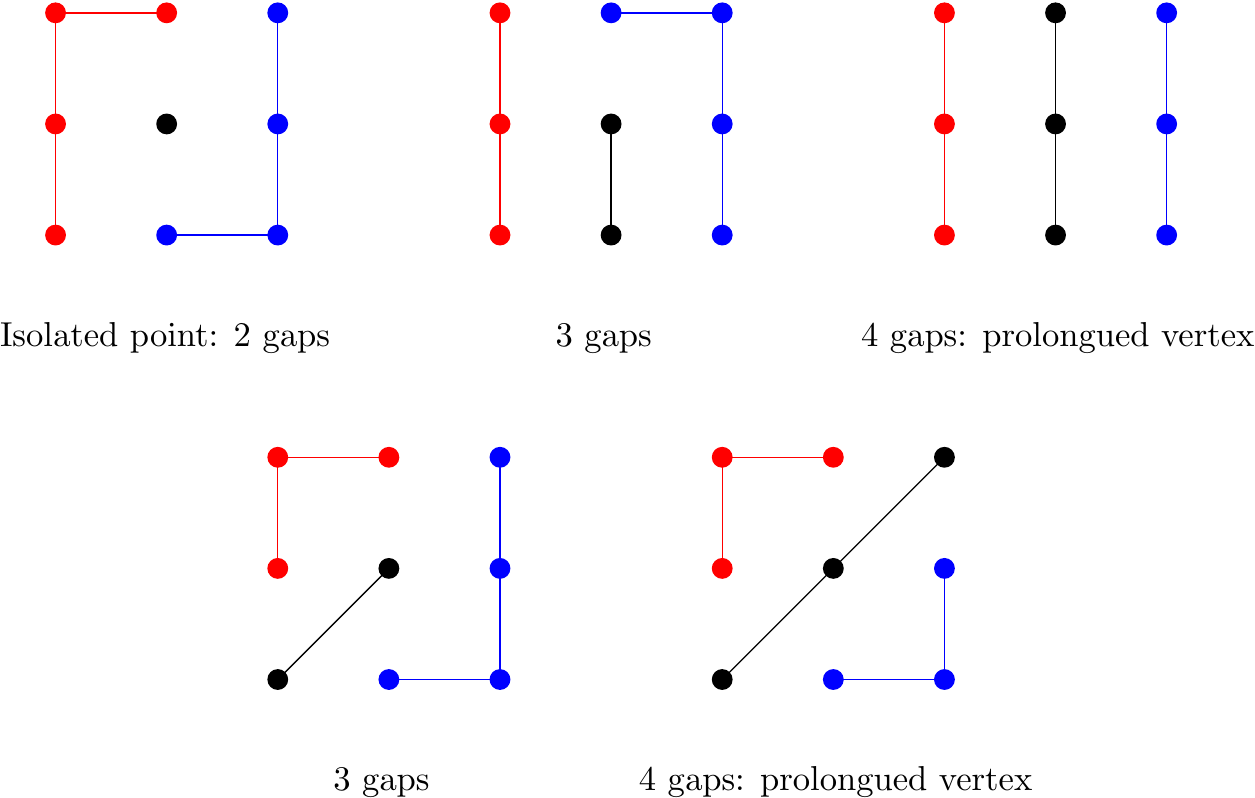}
\caption{Some examples of bulk pathologies: the isolated point, being linked
  to a two-gap frame is automatically not considered as a vertex. The other
  configurations are well handled by the set of rules we prescribed. Indeed,
  for prolongued vertices, the determination of neighbours is always correct.}
\label{fig:bulkPathologies}
\end{figure}

Although we cannot provide a formal proof, we are
  confident that this set of rules correctly handles
all pathological colourings. We can classify these pathologies as
  in \emph{frame pathologies}, Figure \ref{fig:framePathologies} and
\emph{bulk pathologies}, Figure \ref{fig:bulkPathologies}. Indeed, the isolated point
is automatically handled as not being a vertex since it has only 2 gaps in its
frame. The other bulk pathologies are linked to \emph{prolongued vertices}:
configurations where the colour passing through $k$ percolates the entire
stencil. Also prolongued vertices are automatically taken care of, since the
topological information they carry is correct.

A full mathematical proof would require to
handle all different topological configurations for the colouring in the basic
stencil.

Another feature automatically follows: since the colour of obstacles
is set equal to
\texttt{MAXLINK}, it is possible to count the number of gaps along the frame
only if in the gap an obstacle site is not involved. Thus, two Voronoi cells
that are separated by an obstacle will not be considered as neighbours in the
triangulation.

Once one has the adjacency matrix, it is possible to \emph{compress} it with a
standard \emph{unique} operation which automatically sorts the values. This is
due to the fact that the adjacency matrix is initialized with the value
\texttt{MAXLINK} which coincides with the obstacle colour and it is larger
than the maximum number of bubbles, \emph{i.e.}, the maximum number of colours.

\subsubsection{Isomorphism and Trigger}
Proving that two graphs are isomorphic is a challenging task. It has
been recently shown that the general problem can be solved in a
quasi-polinomial time \cite{babai15}. We are not going to face the
general problem since we want to execute the procedure to detect the
topological changes (that we briefly call the \textbf{trigger}) alongside
the simulation. Hence, we need to deal with couples of triangulations
to be compared. We name the two Delaunay triangulation graphs\footnote{Here we adopt the usual
  convention for a graph $G=(V,E)$ which is defined by the sets of vertices
  $V$ and of edges $E$.} as
$\mathcal{D}_{\scriptscriptstyle{P}} =
(V_{\scriptscriptstyle{P}},E_{\scriptscriptstyle{P}})$, where the subscript
$P$ stands for \textbf{past}, and $\mathcal{D}_{\scriptscriptstyle{N}} =
(V_{\scriptscriptstyle{N}},E_{\scriptscriptstyle{N}})$, where the subscript
$N$ stands for \textbf{now}. Since the sets of edges $E_{\scriptscriptstyle{P}}$
and $E_{\scriptscriptstyle{N}}$ are implemented in the two adjacency matrices we
will use both terminologies to indicate the same concept. We can stress this
fact by using the notation
$(E_{\scriptscriptstyle{P}})_{\scriptscriptstyle{P}\scriptscriptstyle{P}}$, or
$(E_{\scriptscriptstyle{N}})_{\scriptscriptstyle{N}\scriptscriptstyle{N}}$,
for the adjacency matrices, where the left subscript indicates the set of
labels used to access the rows and the right subscript the
one used to access the columns. Instead of the classic $(0,1)$ format,
we use {\em custom} values for the adjacency matrix: by default all
the entries are set equal to the value
\texttt{MAXLINK} and whenever two vertices are adjacent we write the vertex
label in its entry for a given row. With this choice we can easily sort the
matrix to compress the values.

\begin{figure}[!t]
\centering
\includegraphics[scale=0.9]{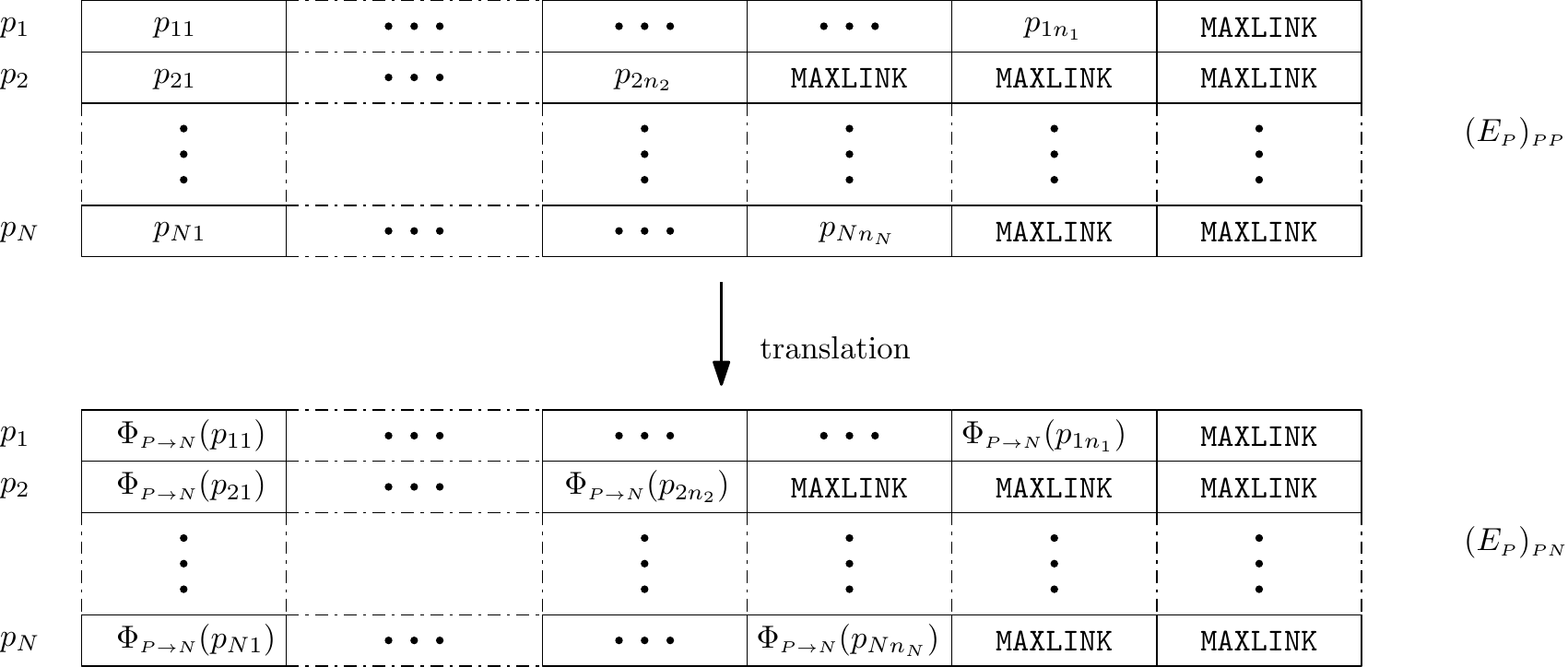}
\caption{Translation process. The adjacency matrix has been reduced to an
  \emph{adjacency list} for which one has a maximum number of entries. We
  label the nodes of the \emph{past} configuration as $p_k$ so that the set of
  vertices can be written as $V_{\scriptscriptstyle{P}} = \{p_k\}$. Then for
  $(E_{\scriptscriptstyle{P}})_{\scriptscriptstyle{P}\scriptscriptstyle{P}}$ the
  row corresponding to the label $p_i$ contains the labels of all the
  sites $p_{ij} \in V_{\scriptscriptstyle{P}}$ neighbouring $p_i$ in the triangulation.
  These are sorted so that each
  row terminates with \texttt{MAXLINK}. We then transform each entry of the
  adjacency list in order to obtain
  $(E_{\scriptscriptstyle{P}})_{\scriptscriptstyle{P}\scriptscriptstyle{N}}$,
  that, in the end, is sorted again in order to check for
  topology changes.}
\label{fig:translation}
\end{figure}

Then, we resort to a kinematic approach: if the
triangulations are obtained from two configurations separated by a
limited time interval, it is possible to obtain the isomorphism by
simply comparing the two sets of centers of mass, \emph{i.e.},
$V_{\scriptscriptstyle{P}}$ and $V_{\scriptscriptstyle{N}}$.
That isomorphism produces an array which is accessed by the labels\footnote{Since each bubble is identified by its label, we extend the terminology in order to refer to the set of the centers of mass, \emph{i.e.} the vertices of a graph $V$, again with the term \emph{label}.}
  $V_{\scriptscriptstyle{P}}$ and
contains the labels $V_{\scriptscriptstyle{N}}$. We can use for the
isomorphism the symbolic notation
$\Phi_{\scriptscriptstyle{P}\to\scriptscriptstyle{N}}$, so that we write
$V_{\scriptscriptstyle{N}} =
\Phi_{\scriptscriptstyle{P}\to\scriptscriptstyle{N}}(V_{\scriptscriptstyle{P}})$.
Then, it is possible to obtain a mixed
definition for the adjacency matrix where the content of the rows is
transformed according to the isomorphism, \emph{e.g.},
$(E_{\scriptscriptstyle{P}})_{\scriptscriptstyle{P}\scriptscriptstyle{N}}$.
In order to use as little memory as
possible, we resort exactly to such a mixed quantity: we tranform the entries
of the past adjacency matrix to the labels $V_{\scriptscriptstyle{N}}$ while keeping the row
indices to be accessed in the labels $V_{\scriptscriptstyle{P}}$. Thus we
obtain the quantity
$(E_{\scriptscriptstyle{P}})_{\scriptscriptstyle{P}\scriptscriptstyle{N}}$. We refer to this
process as \emph{translation} (see Figure \ref{fig:translation}).

After the translation of the past adjacency matrix, we sort both the
$(E_{\scriptscriptstyle{P}})_{\scriptscriptstyle{P}\scriptscriptstyle{N}}$ and
$(E_{\scriptscriptstyle{N}})_{\scriptscriptstyle{N}\scriptscriptstyle{N}}$ so
we can treat them in a \emph{compressed} form. Each compressed adjacency list has a
maximum length, \texttt{linksCompressedSizePast} or \texttt{linksCompressedSizeNow}. The comparison is
performed as follows: each thread accesses a given row of
$(E_{\scriptscriptstyle{P}})_{\scriptscriptstyle{P}\scriptscriptstyle{N}}$ and its
isomorphic one in
$(E_{\scriptscriptstyle{N}})_{\scriptscriptstyle{N}\scriptscriptstyle{N}}$ and
look for two different entries. Given the sorting, this check finds
the vertices involved in the topological change. We keep track of these
changes both with an array named \texttt{local\_trigger} taking
the value 1 for the active indices and 0 for the inactive ones and
with a single variable, \texttt{global\_trigger}, which simply indicates whether
or not a topology change event happened.

However, one needs to consider that \emph{locally} the number of
bubbles can change. For instance, this can happen in confocal microscopy when
the bubble moves away from the focal plane. It is then important to handle
such cases: whenever a bubble disappears the isomorphism is not injective,
\emph{i.e.} two labels of $\mathcal{D}_{\scriptscriptstyle{P}}$ point to a
single one of $\mathcal{D}_{\scriptscriptstyle{N}}$; whenever a bubble appears,
its label in $\mathcal{D}_{\scriptscriptstyle{N}}$ is not matched by any in
$\mathcal{D}_{\scriptscriptstyle{P}}$. It is then possible to isolate such
points and exclude them from the analysis.

\subsubsection{Extracting Topology Change Events}
Since the local trigger indicates which are the active sites, it is necessary
one more step to extract the couples of vertices involved in the topology change event. The entire procedure looks like a \emph{reduction} operation.

As a first step, we count and write on a separate array the {\em triggering} indices. Those are written using the $V_{\scriptscriptstyle{P}}$ labels in the \texttt{eventsIndices} array and their count in \texttt{eventsNumber}.

Then one looks for the \emph{dynamic} indices, \emph{i.e.}, those whose
coordination changes from $\mathcal{D}_{\scriptscriptstyle{P}}$ to
$\mathcal{D}_{\scriptscriptstyle{N}}$. There are two types of such events:
\begin{itemize}
\item those involved in an event that \emph{preserves} the total coordination number
\item those involved in an event that \emph{does not preserve} the total coordination number.
\end{itemize}

The algorithm works as follows: let us consider the triggering indices \texttt{eventsIndices} of the \emph{past} configuration, we need a triple nested
loop, the first (indexed by \texttt{i}) and the second cycling on \texttt{eventsNumber} and
the third on \texttt{linksCompressedSizePast}, then one compares the
content of the adjacency list of an active index with the value of
the array \texttt{eventsIndices} translated to the $V_{\scriptscriptstyle{N}}$ labels
through the isomorphism array \texttt{h\_iso}, representing $\Phi_{\scriptscriptstyle{P}\to\scriptscriptstyle{N}}$. If
  there is a match, the corresponding entry \texttt{eventsCount[i]} is incremented. The
  procedure is repeated for the active indices of the two triangulations. In this way one counts how many active sites are connected to a given active site.

\begin{figure}[!t]
\centering
\includegraphics[scale=1.0]{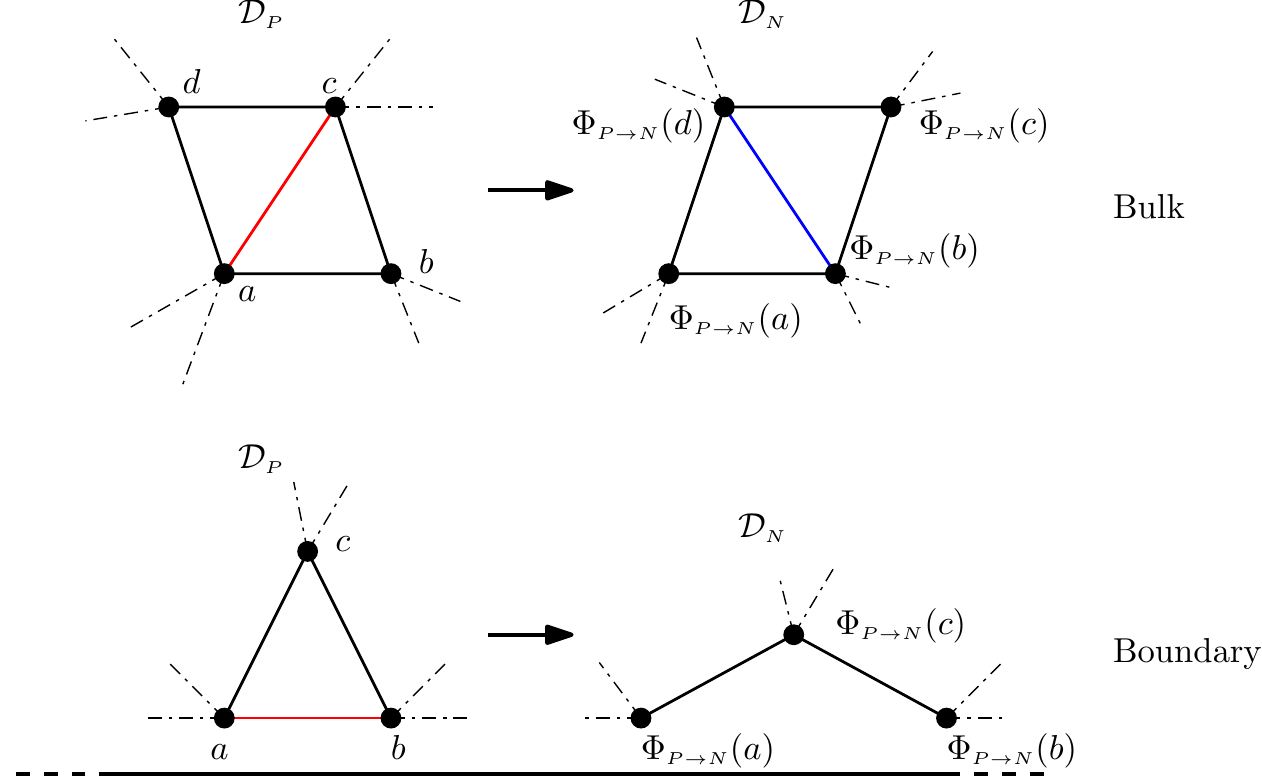}
\caption{Top panel: bulk plastic event conserving the total coordination
  number of the points involved, where the red edge is the
  \emph{breaking link} whereas the blue edge is the \emph{arising link}. Bottom
  panel: boundary plastic event not conserving the total coordination number
  where we have only a \emph{breaking link}.}
\label{fig:T1BulkBoundary}
\end{figure}

For the simplest case, the so-called $T1$ event (see Figure \ref{fig:T1BulkBoundary}), at least four
points are involved in the topology change event.
Hence, in a single
configuration an active index has to be pointed to by, at least, three
other active indices. This is what happens in the first case, for which the total number of links is conserved, and the event must be localized in the bulk of the system.

\begin{figure}
\centering
\includegraphics[scale=0.9]{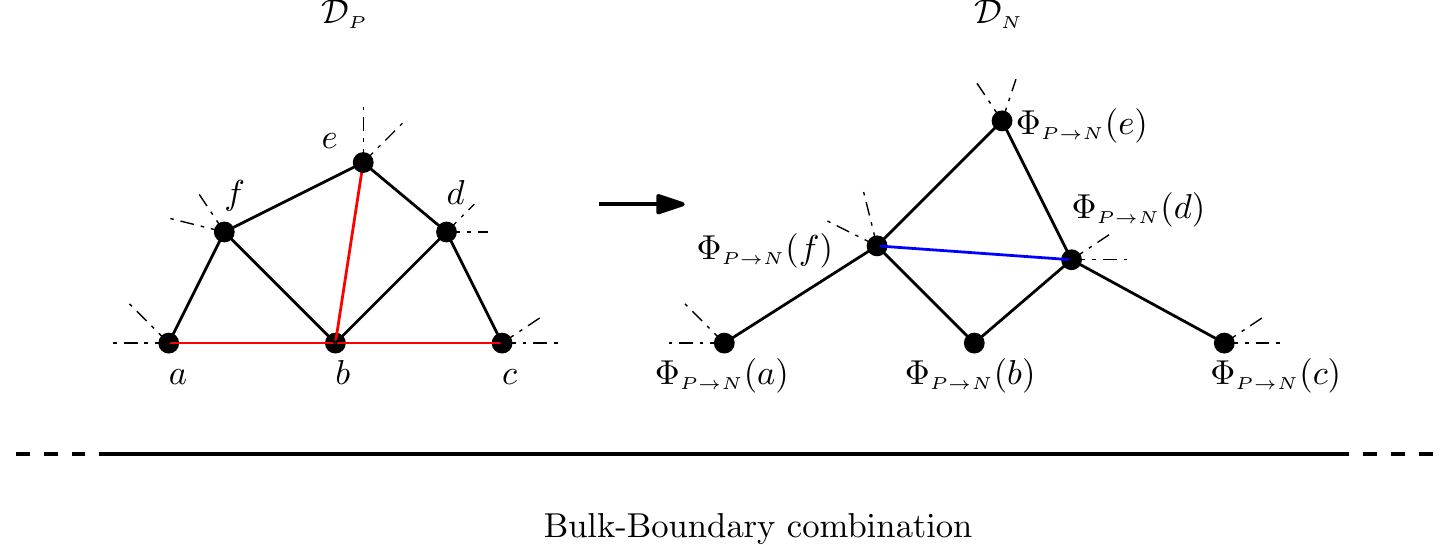}
\caption{Plastic event involving at the same time bulk and boundary
  points. The active boundary point $b$ is connected to other two active
  boundary points $a$ and $c$ which in turn are only connected to $b$. The
  active bulk point $e$ in turn is connected to three other active points
  which are both in the bulk and on the boundary.}
\label{fig:T1BulkBoundaryCombination}
\end{figure}

A topology change event that does not conserve the total number of links might
happen only on the boundary (see Figure \ref{fig:T1BulkBoundary}) of the
triangulation. This assertion is motivated by the fact that if a link in the
bulk disappears, then the triangulation is lost because a quadrangle
appears. For this kind of events, one expects that the indices involved are
pointed to by, at least, one other active index. Of course, it is possible for
a given point to be involved at the same time in both a bulk and a boundary
event (see Figure \ref{fig:T1BulkBoundaryCombination}).

Following this discussion, it clearly appears that it is necessary to
tell apart \emph{bulk} and \emph{boundary} vertices for
the triangulations. As a first step one notices that the concept of
\emph{boundary} is well-defined for the \emph{edge} connecting two
points. Thus, two points connected by a boundary edge belong to
the boundary. Then, it is possible to distinguish between bulk and
boundary edges by using the following criterion:
\begin{itemize}
\item if two connected points share two neighbours, then the edge is in the bulk;
\item if two connected points share only one neighbour, then the edge is on the boundary.
\end{itemize}

Thus, we know how to distinguish points involved in topology change events both in the bulk or on the boundary
\begin{itemize}
\item \textbf{bulk}: the point has to be connected to, at least, three other active points;
\item \textbf{boundary}: the point has to be connected to, at least, one other active boundary point.
\end{itemize}

After having found the active points, we need to determine those
couples that are connected \emph{only} in one of the two
triangulations. As a first step we find which of the
active points are connected and store them as couples. Then, it is
necessary to discard those couples that are given by a simple
permutation of another couple, similarly to \emph{unique}. Finally, one checks
whether or not the couple is connected in both configurations. If the couple
is connected in only one configuration, then this is dubbed either
\emph{breaking} or \emph{arising} if it is found to be in
$\mathcal{D}_{\scriptscriptstyle{P}}$ or in $\mathcal{D}_{\scriptscriptstyle{N}}$ respectively.

To summarize, in order to determine
topological changes we need first to give a unique label to each bubble and
to compute its center of mass: this has been done slightly modifying an already
existing cluster labelling algorithm used for the Ising model cluster dynamics
\cite{Labeling}. Afterwards, we compute the digital Voronoi diagram and its dual
Delaunay triangulation by using a different technique with respect to \cite{DTGPU}
which automatically handles Euclidean colouring pathologies. Finally, we
compare two different triangulations leveraging a
suitable isomorphism linking the two triangulation graphs. Given the adjacency
list we can check if they differ and extract the location of topological changes.

\section{Detection of  emulsions plastic rearrangements \label{sec:lbglassy}}

In this Section we report some preliminary results obtained using the approach we just described. We work in the framework of simulations of soft-glassy materials by means of a suitable multi-component Lattice Boltzmann model \cite{ourJCP,Soft14,Soft15,RoughnessEpl2016}. Such a model relies on a multi-range technique with long-range repulsive and short-range attractive interactions allowing the appearance of a disjoining pressure between the interfaces of two different droplets, thus inhibiting droplets from merging.

The problem is well suited for a GPU parallelization \cite{PREGPU}. Thus, we introduced the plastic events detection as a module into the existing GPU implementation of the model.

\begin{figure}
\centering
\includegraphics[scale=0.6]{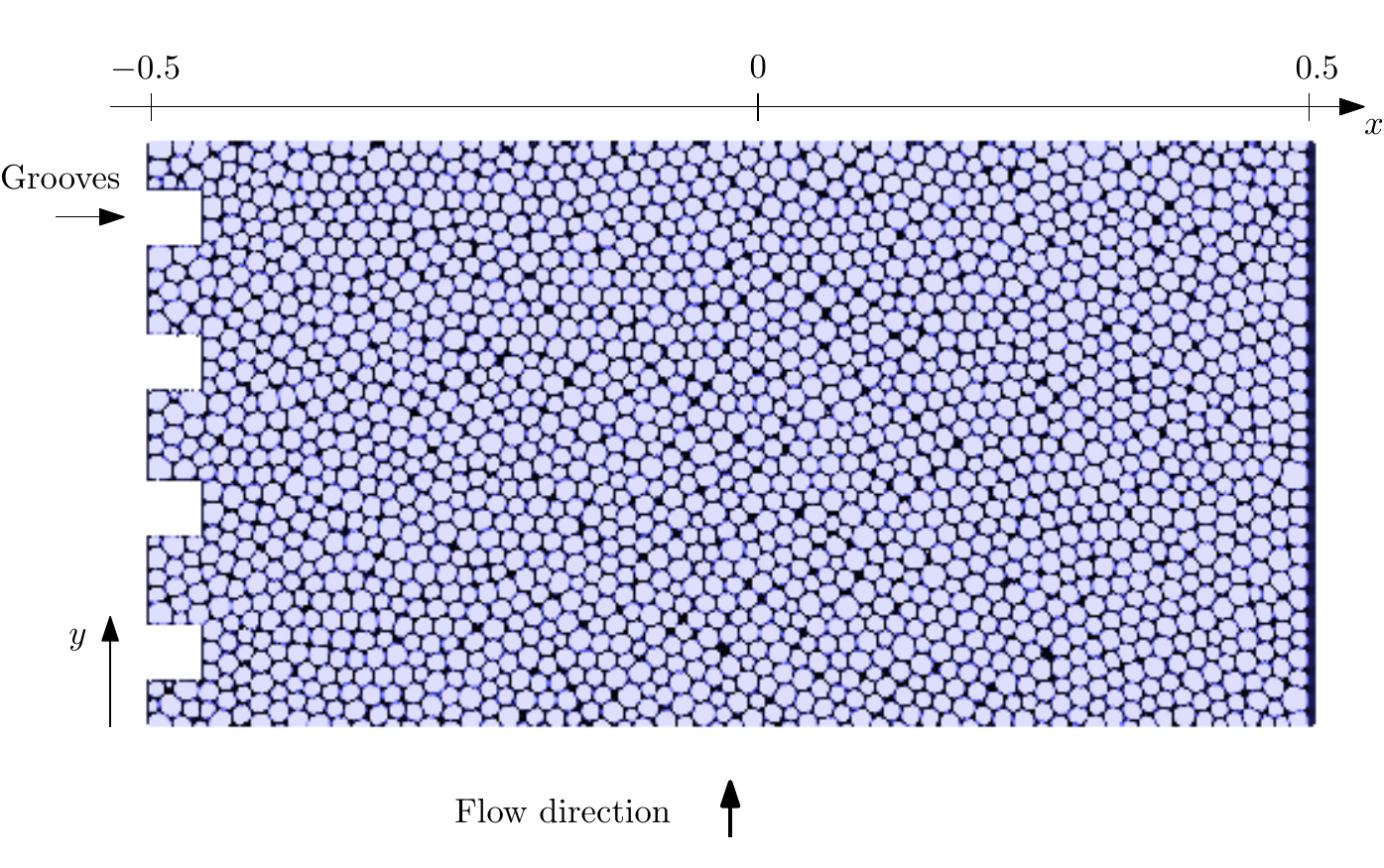}\\
\includegraphics[scale=0.67]{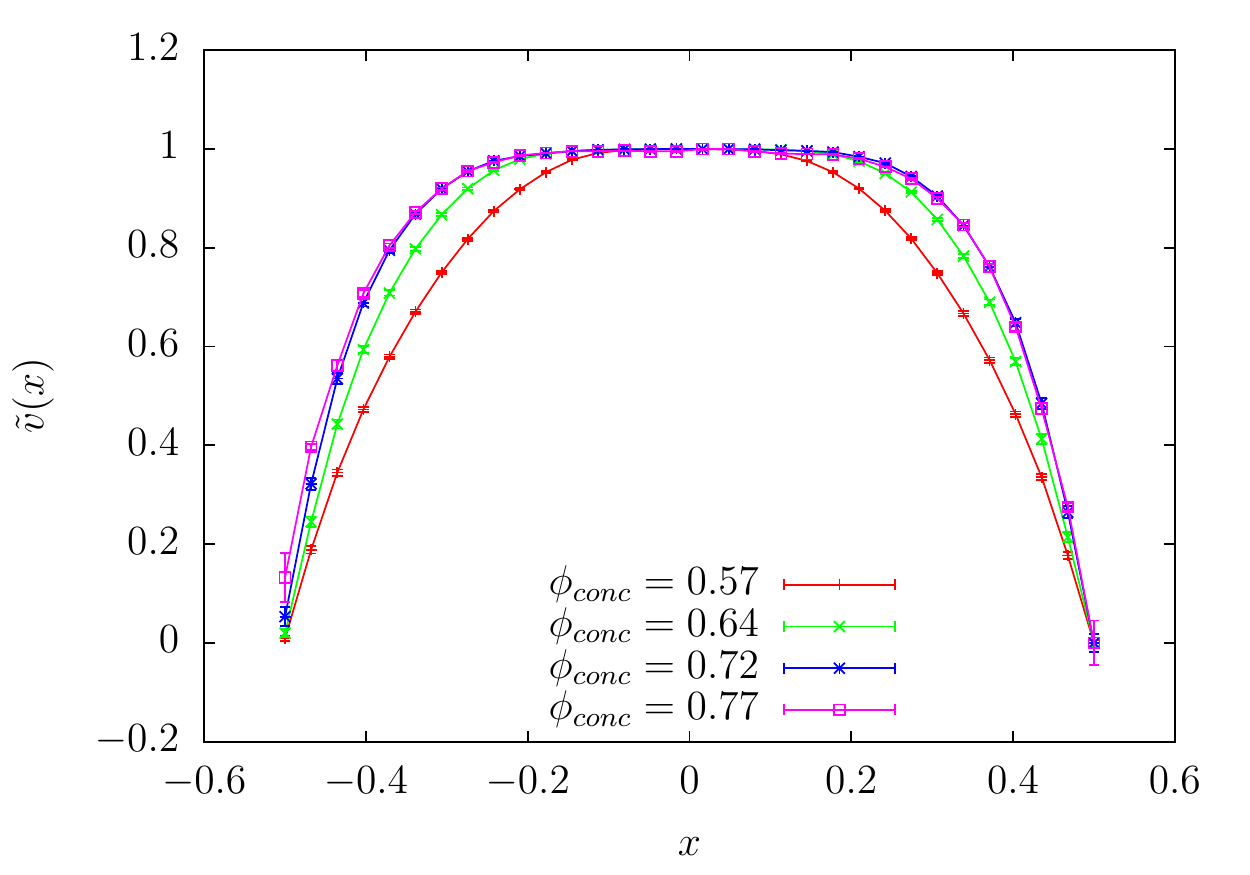}
\caption{Top panel: typical emulsion droplets configuration for a system of sizes $L_x = 2048$ and $L_y = 1024$ with a complex geometry featuring a left wall with equally spaced posts (obstacles) of a given height and width. Applying a pressure gradient we obtain a Poiseuille-like flow. Bottom panel: the velocity profile is averaged along the $y$ direction and normalized to the center channel value. Hence, at the center, \emph{i.e.} $x=0$, we have for the normalized velocity field $\tilde{v}(x=0) = 1$. In the center of the channel, it is possible to see a region of approximately constant velocity (i.e., a ``plug'' flow) which reveals the complex nature of the flow (yield stress fluid \cite{Goyon08}). The higher the packing fraction, the larger is the plug flow region. Results are reported in the wall-to-wall normalized coordinate $x$ with $-1/2<x<1/2$.}
\label{fig:delaunayObstacleT1}
\end{figure}

In the top panel of Figure \ref{fig:delaunayObstacleT1} we show the typical emulsion droplets configuration which we analyze: the domain is resolved with $L_x \times L_y = 2048 \times 1024$ lattice sites, containing roughly $2\cdot 10^3$ droplets with a single droplet diameter resolved with roughly $30$-$35$ lattice sites. The system is highly non-trivial, combining together the complex behaviour of jammed droplets and also a complex geometry, the latter consisting of a confined channel whose left wall is characterized by equally spaced posts (obstacles) of a given height and width. The concentrated emulsion is driven by a pressure gradient applied in the stream-flow direction. We also report four normalized flow velocity profiles $\tilde{v}$ as a function of the normalized channel width $x$, corresponding to four different droplet concentrations $\phi_{conc}$\footnote{The parameter $\phi_{conc}$ expresses the ``concentration'' of the droplets into the continuous solvent. Since we use a diffuse-interface model, we need to set a threshold for the density field in order to determine the inside of a given droplet. Then one simply counts how many lattice sites are above the threshold and takes the ratio with respect to the total number of sites in order to get the concentration: $\phi_{conc} = N_{>thresh}/(L_x L_y)$ . The treshold chosen corresponds to the mean density of the dispersed phase. Notice that for a more quantitative matching with what experimentalists call ``packing fraction'' \cite{Goyon08} one needs to adopt more complicated strategies \cite{RoughnessEpl2016}}. One can see that the velocity profiles develop a more extended plug region in the middle of the channel for higher concentrations, thus stressing the non-Newtonian behaviour of the fluid.

\begin{figure}
\centering
\vcenteredhbox{\includegraphics[scale=0.45]{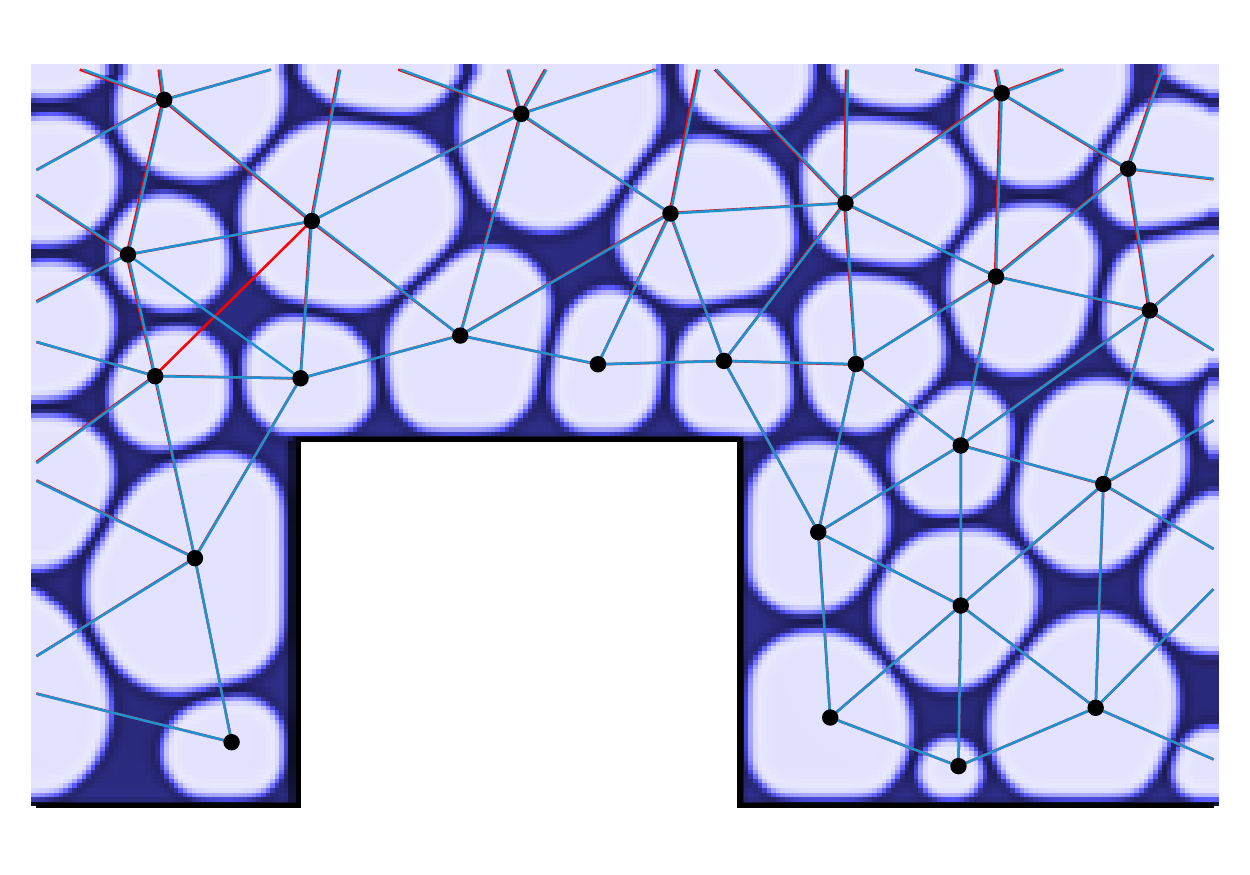}}
\vcenteredhbox{\includegraphics[scale=0.53]{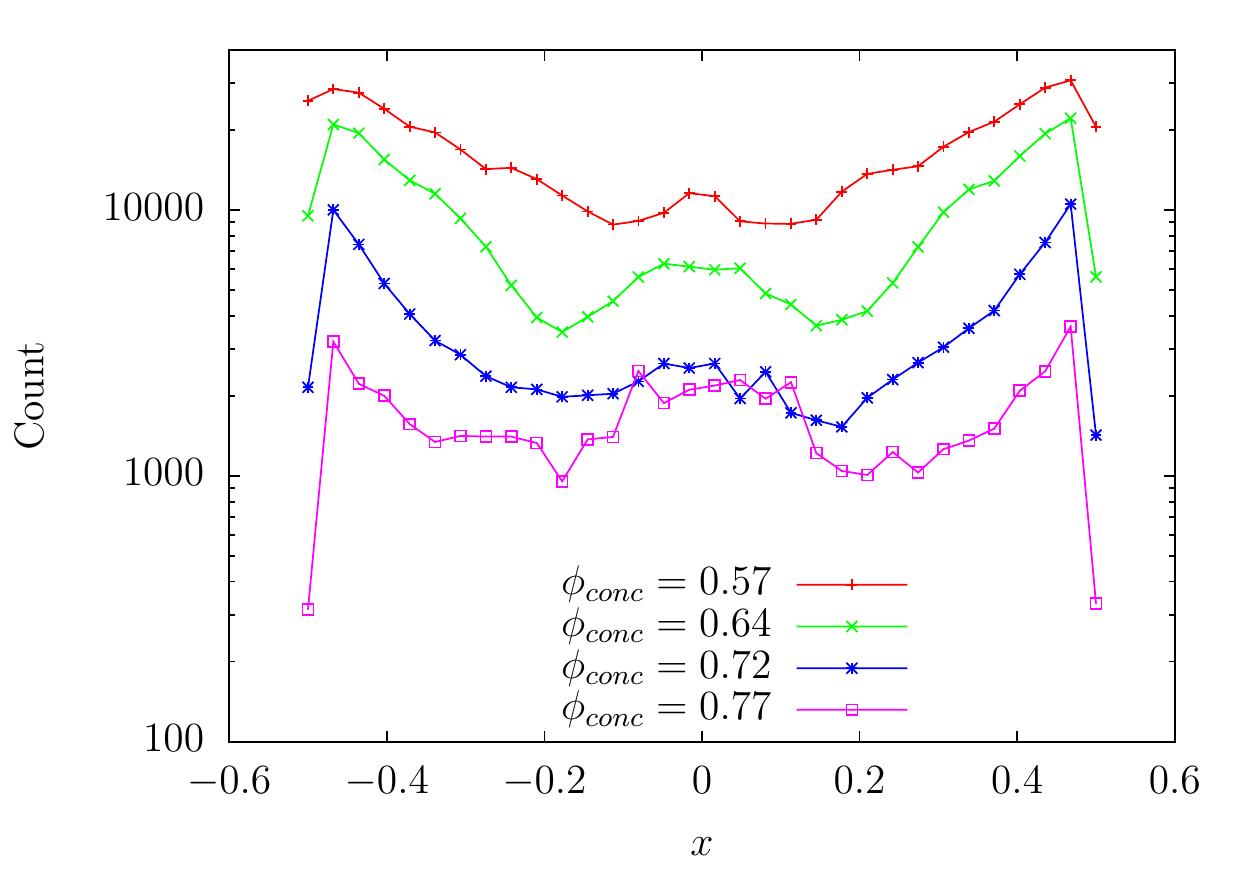}}
\caption{Left panel: comparison between triangulations containing boundary and bulk points. The red link is a \emph{breaking} link whereas the light blue one is an \emph{arising} link. Right panel: total count of plastic events as a function of the normalized coordinate. Results range three orders of magnitude.}
\label{fig:poiseuilleResults}
\end{figure}

The simulation time is $T=4\cdot 10^6$ time steps and Delaunay triangulations were compared every $\Delta t = 100$ time steps. The overhead induced by this analysis on the simulation lies within the typical running time fluctuations. In other words,
using or not the GPU procedure does not affect the total execution time of the simulation, whereas with the previous CPU based solution, we could check for topological changes only every few thousands time steps otherwise the impact on the execution time and the space requirements would have been too high.
To be noticed that the proposed algorithm is capable of detecting plastic activity both in the bulk of the system and within the grooves separating the obstacles of the complex geometry. This analysis also allows to quantitatively highlight features on the orientation of the plastic events, by defining ``breaking'' and ``arising'' links (see left panel of Figure \ref{fig:poiseuilleResults} and also \cite{RoughnessEpl2016}). The right panel of Figure \ref{fig:poiseuilleResults} reports preliminary results for the counting of plastic events as a function of the packing fraction: one can see that for the lower packing fraction, the number of plastic events is the highest and it reduces for increasing packing fractions. When moving from smaller to larger packing fractions, the distribution of plastic events tends to flatten, which may be taken as a signature of the emergence of a solid-like behaviour into the system \cite{Soft14,Soft15}. It is interesting to notice that the number of plastic events vary of three orders of magnitude with a $50\%$ difference for the packing fractions. We can handle such a wide range of variation obtaining a very high number of events detection when needed.

We finally remark that the results presented in this Section are only meant to highlight interesting physical properties of concentrated emulsions that one can study with the procedure on CUDA kernels described in the paper. This surely calls for a more systematic investigation in the coming future: with the tool at hand, one can in fact study correlations of plastic events, their dependence on both the load conditions and the complex geometry, with a reliable statistics. We believe this will bring forth new and stimulating results for the understanding of the physics of yielding fluids in the coming future.

\section{Conclusions and future perspectives \label{sec:conc}}
We presented a software procedure to detect topological changes in
Voronoi diagrams based on a set of CUDA kernels that we are going to release in open source format from \verb|http://twin.iac.rm.cnr.it/dynvorcuda.tgz|.

We tested the procedure during simulations of soft-glassy matter running on GPU with the
goal of detecting plastic events. Previously we had to copy data from GPU to CPU to use
a CPU library for Voronoi tesselation that represented a major bottleneck and a serious
performance limiting factor of the whole simulation process. With the new procedure we expect to be in a position that will allow us to perform a side-by-side comparison with experimental results and we have provided here some related data.

The procedure can be applied to a wide set of scenarios. The first parts can be used
to identify set of points on a lattice and characterize them according to their size (number of points of each set). The only requirement is a criterion to identify a point
as belonging to a set. The other parts can be used to track the dynamics of Voronoi diagrams in a way that does not depend on the procedure by which the diagrams have been generated. We expect to refine and further generalize the procedure. A challenging extension is the development of a multi-GPU implementation that could manage much larger datasets overcoming the limitation imposed by the memory of a single GPU.

The research leading to these results has received funding from the European Research Council under the European Community's Seventh Framework Programme (FP7/2007-2013)/ERC Grant Agreement no. [279004]

\bibliographystyle{elsarticle-num}

\end{document}